\documentclass{iau_FM}
\usepackage{graphicx}

\title[An Overview of Inside-Out Planet Formation] 
{An Overview of Inside-Out Planet Formation}

\author[J. C. Tan et al.]   
{Jonathan C. Tan$^{1}$, Sourav Chatterjee$^2$, Xiao Hu$^3$, Zhaohuan Zhu$^4$
 \and Subhanjoy Mohanty$^5$}

\affiliation{$^1$Depts. of Astronomy \& Physics, University of Florida, Gainesville, FL 32611, USA\\ email: {\tt jctan.astro@gmail.com}\\[\affilskip]
$^2$CIERA, Physics and Astronomy, Northwestern University, Evanston, IL 60208, USA\\[\affilskip]
$^3$Dept. of Astronomy, University of Florida, Gainesville, FL 32611, USA\\[\affilskip]
$^4$Dept. of Astrophysical Sciences, Princeton University, Princeton, NJ 08544, USA\\[\affilskip]
$^5$Dept. of Physics, Imperial College, London, UK
}

\pubyear{2015}
\jname{Astronomy in Focus, Volume 1} 
\editors{Piero Benvenuti, ed.}
\begin{document}

\maketitle

\begin{abstract}
The {\it Kepler}-discovered Systems with Tightly-packed Inner Planets
(STIPs), typically with several planets of Earth to super-Earth masses
on well-aligned, sub-AU orbits may host the most common type of
planets, including habitable planets, in the Galaxy. They pose a great
challenge for planet formation theories, which fall into two broad
classes: (1) formation further out followed by inward migration; (2)
formation {\it in situ}, in the very inner regions of the
protoplanetary disk. We review the pros and cons of these classes,
before focusing on a new theory of sequential {\it in situ} formation
from the inside-out via creation of successive gravitationally
unstable rings fed from a continuous stream of small ($\sim$cm-m size)
``pebbles,'' drifting inward via gas drag. Pebbles first collect at
the pressure trap associated with the transition from a
magnetorotational instability (MRI)-inactive (``dead zone'') region to
an inner, MRI-active zone. A pebble ring builds up that begins to
dominate the local mass surface density of the disk and spawns a
planet. The planet continues to grow, most likely by pebble accretion,
until it becomes massive enough to isolate itself from the accretion
flow via gap opening. This reduces the local gas density near the
planet, leading to enhanced ionization and a retreat of the dead zone
inner boundary. The process repeats with a new pebble ring gathering
at the new pressure maximum associated with this boundary. We discuss
the theory's predictions for planetary masses, relative mass scalings
with orbital radius, and minimum orbital separations, and their
comparison with observed systems. Finally, we discuss open questions,
including potential causes of diversity of planetary system
architectures, i.e., STIPs versus Solar System analogs.

\keywords{formation --- planets and satellites, protoplanetary disks}
\end{abstract}

\firstsection 
\section{Introduction}

Thousands of exoplanets have been discovered, especially by NASA's
{\it Kepler} mission (e.g., Mullally et al. 2015), and most are in
systems that are quite different from our own Solar System. In
particular, a large percentage ($\gtrsim 30\%$) of low-mass stars are
now thought to host Systems with Tightly-packed Inner Planets
(STIPs). These usually have 3 or more detected planets of radii
$\sim1-10\:R_\oplus$ on orbital periods from $\sim 1$ to 100 days with
a peak at $\sim 10$ to 20 days, i.e., orbital radii of $\sim 0.1$~AU
(e.g., Fang \& Margot 2012).
Also, the systems are ``tightly-packed,'' i.e., with period ratios
near 1.5 to 3, equivalent to separations of $\sim 10$ to several tens
of Hill radii, but are not on the verge of instability (as expected,
since they are generally billions of years old). The period ratios are
mostly non-resonant, with only $\sim 10\%$ piled-up just wide of first
order resonances (mostly 2:1 and 3:2). They have a low dispersion in
orbital inclination angles ($\lesssim 3^\circ$). From the small subset
of planets with dynamical mass measurements, we know that there is a
wide range of mean densities of a factor of several, which indicates
that some STIPs planets have accreted a H/He atmosphere that is a
few \% of the total mass. STIPs may host the most common kind of
planet in the Universe and the most common type of habitable
environments, which would be in STIPs around K and M main sequence
stars.

The first theoretical scenario that has been proposed to explain STIPs
involves formation of planets in the outer disk via the Core Accretion
paradigm, followed by migration to the inner region (e.g., McNeil \&
Nelson 2010; Kley \& Nelson 2012). Note that these models have
generally assumed protoplanets are able to form from the outer disk,
but have not explicitly model this step (c.f., Lambrechts \& Johansen
2014; Levison et al. 2015; Bitsch et al. 2015), i.e., their initial
conditions already involved massive protoplanets that are placed at
quite arbitrary locations.

These models have faced several problems in reproducing the observed
exoplanet systems. For example, McNeil \& Nelson (2010) found it
difficult to concentrate planets close to their host star to the
degree observed in STIPs. Another major problem is that planets
undergoing significant migration tend to become trapped in low-order
mean motion resonances, which, as discussed above, are not a
particular feature of the observed systems. This has then motivated
other work to identify potential mechanisms of either reducing the
efficiency of resonant trapping (Goldreich \& Schlichting 2014) or to
later move them out of resonance (e.g., Lithwick \& Wu 2012; Rein
2012; Batygin \& Morbidelli 2013; Chatterjee \& Ford 2015).

As a very different alternative, {\it in situ} formation of the STIPs has
been discussed by Chiang \& Laughlin (2013) and modeled by Hansen \&
Murray (2012, 2013). However, this modeling again involves starting
with a population of protoplanets (some as massive as 6~$M_\oplus$)
that are initially distributed in a very concentrated region inside
about 1~AU. After 10~Myr of collisional N-body evolution, Hansen \&
Murray found that oligarchic growth had led to planetary architectures
similar to those of STIPs, including a relatively flat distribution of
planetary masses with orbital radius. However, Ogihara et al. (2015)'s
study, which is similar but also includes the effect of gas and
resulting protoplanetary migration, leads to systems with planet
masses that decline steeply with orbital radius, which are very
different from the observed STIPs, and thus argues against this
{\it in situ} oligarchic growth phase.

\section{Inside-Out Planet Formation - Theoretical Summary}

\begin{figure}[t]
\vspace*{-0.25 cm}
\begin{center}
 \includegraphics[width=5in]{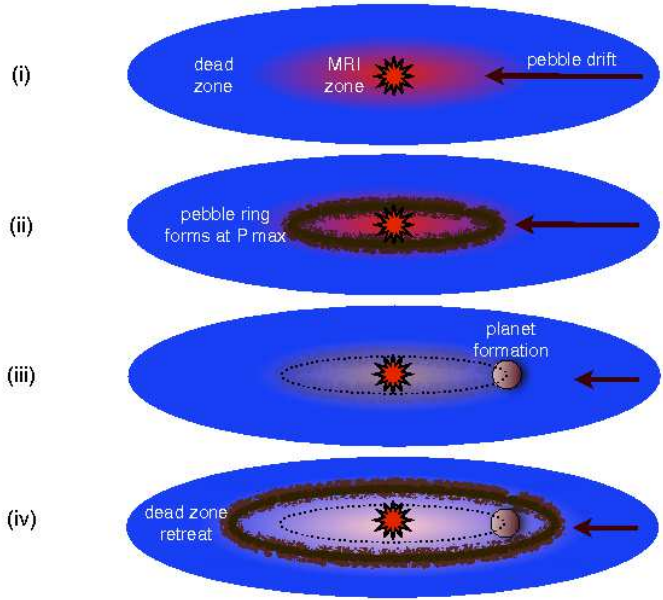} 
\vspace*{-1.155 cm}
 \caption{
Schematic overview of Inside-Out Planet Formation (CT14). {\bf (i)
  Pebble formation and drift to the inner disk.} Pebbles form via dust
coagulation in the protoplanetary disk. Those with $\sim$cm to m sizes
attain high radial drift velocities and quickly reach the dead zone
inner boundary (DZIB), where they become trapped at the pressure
maximum. {\bf (ii) Pebble ring formation.} A ring of pebbles gradually
builds up over a timescale set by the pebble formation and supply rate
from the outer disk. {\bf (iii) Planet formation and gap opening.} A
planet forms from the pebble ring and continues to grow by pebble
accretion until it becomes massive enough to open a gap. This shuts
off pebble accretion and may also lead to reduction in gas supply to
the inner disk, which would then dissipate by viscous clearing. {\bf
  (iv) Dead zone retreat and subsequent pebble ring and planet
  formation.} Gap opening and potential viscous clearing of the inner
disk lead to lower densities and greater penetration of X-ray photons
from the protostar to the disk midplane, increasing its ionization
fraction and thus activating the MRI. The inactive dead-zone retreats,
along with the pressure maximum associated with its inner boundary. A
new pebble ring starts to form at this location that forms a new
planet. This cycle repeats, leading to sequential formation of a
planetary system from the inside-out.}
   \label{fig1}
\end{center}
\end{figure}

An overview of the Inside-Out Planet Formation (IOPF) model
(Chatterjee \& Tan 2014, hereafter CT14 or Paper I) is shown in
Fig.\,\ref{fig1}. The first basic assumption is that there is
efficient supply of ``pebbles'' drifting radially inwards to
$\sim$0.1~AU from the outer disk. This radial drift is a well-known
effect due to gas drag in regions where the gas disk derives some
support from a radially decreasing pressure gradient causing its
orbital speeds to be slightly sub-Keplerian (Weidenschilling
1977). Indeed this drift is so strong that it has long been recognized
as a major problem for planetesimal formation, which is part of the
so-called ``meter-sized barrier.'' The radial drift of pebbles is
assumed to be stopped at the local pressure maximum associated with
the dead zone inner boundary (DZIB), i.e., where gas and pebbles both
orbit at the Keplerian speed so that there is no headwind gas drag
experienced by the pebbles. The location of this DZIB is assumed to be
set by when gas temperatures reaches about
1,200~K, allowing thermal ionization of alkali metals Na and K
(Umebayashi \& Nakano 1988). These species should provide enough
ionization to allow the magneto-rotational instability (MRI) (Balbus
\& Hawley 1991) to operate, which increases the disk's viscosity and
so leads to reduced surface densities, volume densities and pressures
compared to at the DZIB. A pebble ring then builds up at the DZIB,
which can come to dominate the local mass surface density. A planet
forms from this ring. The protoplanet grows without suffering
significant migration. The next crucial stage is when the planet,
which has been growing by pebble accretion, becomes massive enough to
open a (potentially quite shallow) ``gap'' in the disk that is
sufficient to move the local pressure maximum away from the planet,
thus shutting off pebble accretion. At the same time, the reduction in
gas density around the planet leads to increased ionization, perhaps
also due to increased X-ray penetration from the protostar, activating
the MRI and causing the DZIB to retreat outwards. This retreat can be
self-propagating since increasing viscosity in the boundary region
leads to further reductions in densities. However, this processes
stabilizes relatively quickly and a new pebble ring begins to form at
the pressure maximum at the retreated DZIB. This location will be at
least several Hill radii from where the first planet formed, but could
be significantly further away. The entire process repeats leading to
the sequential formation of a compact, well-aligned planetary system
from the inside-out.

In order to make quantitative estimates, CT14 adopted the Shakura \&
Sunyaev (1973) ``$\alpha$-disk'' model framework for the structure of
a steady, active accretion disk, i.e., in which the heating is
dominated by accretion. Typical observed accretion rates of T-Tauri
stars (Alcala et al. 2014) and stars with transition disks (Manara et
al. 2014) are $\sim 10^{-9}\:M_\odot\:{\rm yr^{-1}}$, with a
dispersion of about a factor of 100.
Transition disks in which there are gaps and holes in the very inner
disk dust distribution inside $\sim 1$~AU may be particularly relevant
for the IOPF model.  Thus CT14 adopted $\dot{m}=10^{-9}\:M_\odot\:{\rm
  yr^{-1}}$ as a fiducial value, i.e., $\dot{m}_{-9}=1$, but consider
potential variations of $\dot{m}_{-9}=0.1$ to 10. For simplicity, CT14
also adopted a fixed opacity of 10~$\rm cm^2\:g^{-1}$, i.e.,
$\kappa_{10}=1$, which is a typical value expected in inner
protoplanetary disks (e.g., Zhu et al. 2009).
The value of the $\alpha$ viscosity parameter in protoplanetary disks
is quite uncertain. In DZIB regions, the simulations of Dzyurkevich et
al. (2010) find effective viscosities equivalent to $\alpha\sim
10^{-4}$ to $10^{-3}$, partly set by the propagation of turbulence
outwards from the MRI-active region. CT14, with a focus on disk
midplane conditions, adopted $\alpha=10^{-3}$ as a fiducial value in
the dead zone region (but we will see below that moderately smaller
values may be preferred). In the MRI-active region, $\alpha$ is
assumed to rise to much larger values $\sim 10^{-2}$ or more.

In the context of this accretion disk model, CT14 showed that the
radial drift time of pebbles from the outer to inner disk was very
short compared to expected disk lifetimes. Hu, Tan \& Chatterjee
(2014) presented more detailed calculations, including Stokes-limited
pebble growth via sweep-up of small grains, finding that initially
1~mm-radius pebbles would reach the inner disk after only 2,000 or
40,000~yr if starting from 10 or 100~AU, respectively (this assumes
the dead zone value of $\alpha$ extends to these scales). However, a
quantitative estimate of the pebble production rate and thus the
overall mass flux in pebbles to the inner disk has not yet been made
for these models. Still, observations of disks are beginning to reveal
both radial concentrations of dust with respect to gas (e.g, de
Gregorio-Monsalvo et al. 2013) and increasing grain sizes in the inner
regions (e.g., P\'erez et al. 2012; Trotta et al. 2013), so a large
mass flux of pebbles to inner disks remains a distinct and even likely
possibility.

CT14 evaluated the location of the DZIB by the condition that disk
midplane temperature reaches 1,200~K, finding a radius $r_{\rm
  1200K}=0.1\phi_{\rm
  DZIB,0.1AU}\kappa_{10}^{2/9}\alpha_{-3}^{-2/9}m_{*,1}^{1/3}
\dot{m}_{-9}^{4/9}$~AU, with $\phi_{\rm DZIB,0.1AU} = 1.8$, i.e., a
fiducial location of 0.18~AU (around a star of 1~$M_\odot$, i.e.,
$m_{*,1}=1$).  Note that the location of the DZIB increases for
larger accretion rates. Hu et al. (2015) (Paper III) revisited the
disk structure equations and adopted a slightly different choice for
normalization of the vertical optical depth equation (or equivalently
the definition of midplane conditions), which leads to an estimate of
$r_{\rm 1200K}=0.13$~AU, i.e., $\phi_{\rm DZIB,0.1AU}=1.3$. Mohanty \&
Tan (in prep.) considered the structure of a fully self-consistent
MRI-active inner disk, finding $\alpha$ decreased rapidly due to Ohmic
resistivity at a radius of $\sim 0.1$ to 0.2~AU.

CT14 discussed various potential mass scales of planet formation from
the pebble ring, including the Toomre mass from a gravitationally
unstable ring ($\sim10^{-3}\:M_\oplus$ in the fiducial case) and the
Toomre Ring mass (fiducial value of $\sim1\:M_\oplus$). However, the
most important mass scale is identified as being the gap-opening mass
(Lin \& Papaloizou 1993), $M_G = \phi_G 40 \nu m_*/(r^2\Omega_K)\simeq
5.59 \phi_{G,0.5}\kappa_{10}^{1/5}\alpha_{-3}^{4/5}
m_{*,1}^{3/10}\dot{m}_{-9}^{2/5}r_{\rm 0.1AU}^{1/10}\:M_\oplus$, which
is derived by considering the competition of the planet's gravity with
the viscosity of the gas. Here the overall normalization, including
choice of $\phi_G=0.5$, is based on the numerical simulations of Paper
III: at this mass scale the response of the disk to the presence of
the planet leads to the pressure maximum being displaced outwards by
about 5~$R_H$.


The mass scale for gap opening at the location of the DZIB set by
midplane temperature of 1,200~K, which would be the mass of innermost,
``Vulcan'' planets in the IOPF model, has the following dependencies
(Chatterjee \& Tan 2015, hereafter CT15 or Paper II):
$M_{p,1}=M_G(r_{\rm 1200K})=5.59 \phi_{G,0.5} \phi_{\rm
  DZIB,0.1AU}^{-9/10} \alpha_{-3} r_{\rm 0.1AU}\:M_\oplus$ (note the
normalization here follows the Paper III disk model and is a factor of
0.745 smaller than in Paper II). This prediction is that inner planet
mass scales linearly with orbital radius, does not depend on $\kappa$
or $m_*$, but does depend on the value of $\alpha$ in the DZIB region.

The question of the potential migration of protoplanets as they are
forming and opening gaps has been studied in Paper III, where we find
that from 0.1 to 1~$M_G$, protoplanets are trapped very close to their
formation location set by the initial pressure maximum (and associated
gas surface density maximum) at the DZIB.

Subsequent retreat of the DZIB due to gap opening and increased X-ray
penetration has been studied with simple, heuristic models in Paper
III. Important parameters include the penetration depth of X-rays
through the gas disk that allows MRI activation and the width the
transition zone from the MRI-active to inactive regions. Paper III
presented simple example models of this process that could lead to
DZIB (and thus pressure maximum) retreat by several tens of Hill radii
of the already-formed planet.

\section{Inside-Out Planet Formation - Observational Summary}

\begin{figure}[t]
\begin{center}
 \includegraphics[width=5in]{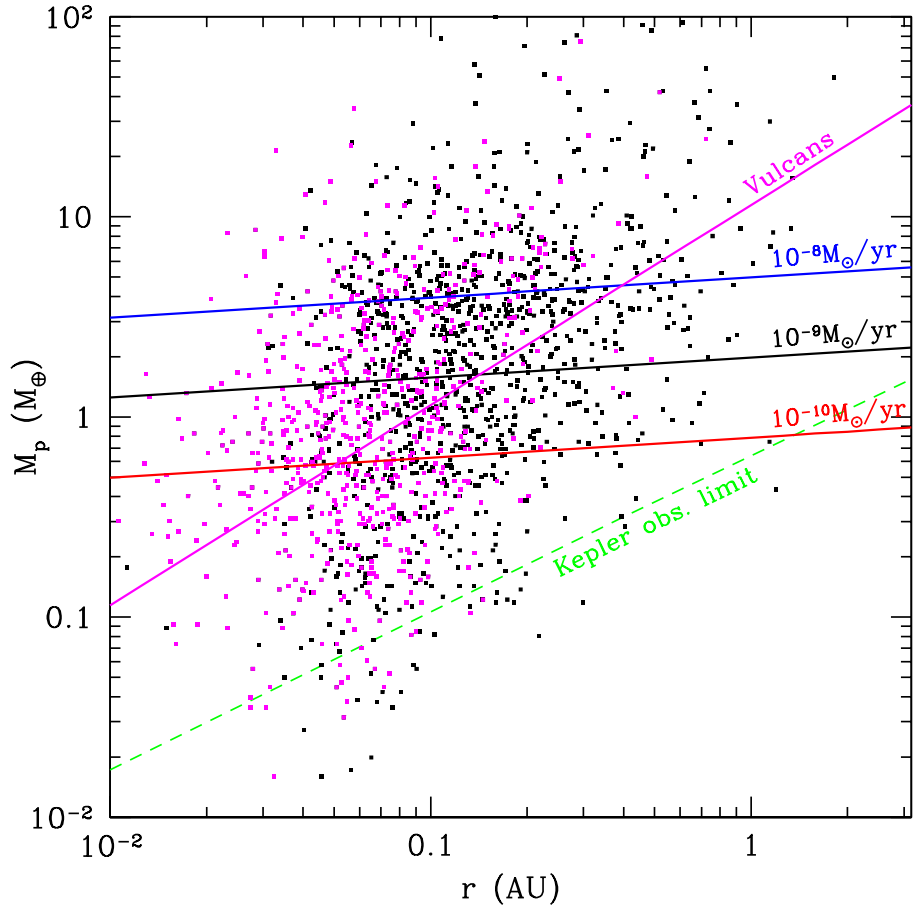} 
\vspace*{-0.3cm}
 \caption{
Planet mass versus orbital radius for 1,656 STIPs planets, with
664 innermost, ``Vulcan'' planets shown in magenta. The gap opening mass,
$M_G$, for disks with $\alpha=2.05\times10^{-4}$ and
$\dot{m}_*=10^{-10}, 10^{-9}, 10^{-8}\:M_\odot\:{\rm yr}^{-1}$ are
shown with the solid red, black and blue lines, respectively. The
solid magenta line shows $M_G(r=r_{\rm 1200K})$ for these disks (with
$\phi_{\rm DZIB,0.1AU}=1$), which is the Inside-Out Planet Formation
model prediction for Vulcan planet masses. The dashed green line shows
an approximate estimate for {\it Kepler's} detection limit (S/N=7) for
the median $K_p=14.5$ host star and the CT15 PL3 mass-size relation.}
\label{fig2}
\end{center}
\end{figure}

CT14 noted that if the dead zone inner boundary is set due to thermal
ionization of alkali metals at $\sim 1,200$~K, then its expected
location in disks with accretion rates of $\sim 10^{-9}\:M_\odot\:{\rm
  yr}^{-1}$ (i.e., similar to those of observed stars with transition
disks) is estimated to be $\sim 0.1$~AU. This is very similar to the
sizes of the observed orbits of the STIPs planets. The expected mass
scale for gas gap opening is several Earth masses, assuming
$\alpha\sim10^{-3}$, which is again similar to the STIPs planet
masses. However, it should be noted that most of these mass estimates
are quite uncertain, since they are based on an assumed mass (or
density) versus size relation (e.g., Lissauer et al. 2011) and it is
clear from the planets with dynamical mass measurements that at a
given mass there is actually a wide range of densities of a factor of
about 5 (CT15). Fig.~\ref{fig2} plots the masses and orbital radii of
the most recent census of STIPs planets, where mass has been estimated
from the piecewise power law fit to the mass-size relation of STIPs
with dynamical mass measurements (PL3 of CT15). The analytic values
for gap opening masses assume $\alpha_{-3}=0.205$, which is based on a
comparison of only the innermost, Vulcan planets (CT15), discussed
below. We see in Fig.~\ref{fig2} that the mass scales and orbital
locations of STIPs planets are consistent with gap opening masses near
DZIBs in disks with typical observed accretion rates.

CT14 also examined the dependence of planet mass with orbital
radius. Around a given star and for a constant accretion rate, the gap
opening mass is expected to scale as $M_G\propto r^{k_M}$ with
$k_M=0.1$, i.e., a relatively flat scaling. Examining the 4, 5 and
6-planet systems known at the time, CT14 found that masses scaled as
power laws with orbital radius with indices of $k_M=0.92\pm0.63$,
$0.78\pm 0.64$ and 0.50, respectively, with the quoted uncertainty
being the dispersion in the samples. For the handful of systems that
had dynamical mass measurements of their planets, CT14 found
$k_M=1.0\pm2.1$ (average of 6 systems) and $k_M=0.47\pm2.7$ (average
of all adjacent planet pairs). In summary, the data are consistent
with the scaling predicted by the gap opening mass, although there is
a hint that observed planet masses increase more steeply with orbital
radius. However, for the trends derived from the STIPs that are
lacking dynamical mass measurements, the results are strongly
influenced by the choice of planet density with orbital radius: a
systematic trend of denser inner planets, due either to pebble
composition during formation or effects of subsequent evaporation,
would tend to lower the values of $k_M$.

Papers I \& III examined the orbital spacings between adjacent planet
pairs, normalized by the Hill radius of the innermost planet,
$R_{H,i}$, i.e., $\phi_{\Delta r,i}\equiv \Delta r_i/R_{H,i}$, where
$\Delta r_i = r_{i+1}-r_i$. In the IOPF model we expect the first gap
opening event and potential clearing of the inner disk to lead to
relatively larger DZIB retreat, perhaps due to increased X-ray
penetration from the protostar. However, subsequent planet formation
events would have a more incremental effect on the disk structure and
may be expected to lead to more modest retreats and thus smaller
normalized orbital separations. In Papers I \& III, the latter using
improved STIPs planet mass estimates from Paper II, we do find
statistically significant differences in the distributions of
$\phi_{\Delta r,1}$ compared to those of the other planet pairs (which
themselves have indistinguishable distributions). The distribution of
$\phi_{\Delta r,1}$ peaks at larger values $\sim20$ to 40, while
$\phi_{\Delta r,2}$, $\phi_{\Delta r,3}$ \& $\phi_{\Delta r,4}$ peak
at $\lesssim20$. For example, in systems with $\geq3$ planets, the
probability that the observed distributions of $\phi_{\Delta r,1}$
\& $\phi_{\Delta r,2}$ are drawn from the same underlying
distribution is only $9\times 10^{-4}$, and restricting to systems
with $\geq4$ planets the probabilities are only $\sim 10^{-6}$ that
$\phi_{\Delta r,1}$ has the same distribution as $\phi_{\Delta r,2}$ or
$\phi_{\Delta r,3}$ (Paper III). These differences are interesting
observational results that, in the context of IOPF, impose constraints
on models of DZIB retreat.

CT15 tested the predicted mass versus orbital radius scaling of
innermost, Vulcan planets, which is shown by the magenta line in
Fig.~\ref{fig2}. For each host star of a detected Vulcan, a planet
with the appropriate gap opening mass, $M_G(r_{\rm 1200K})$, with
$r_{\rm 1200K}$ set equal to the current observed orbital radius, was
modeled. These planets were given densities (and thus sizes) randomly
sampled from the distribution functions fitted to observed STIPs
planets with dynamical mass measurements. It was then checked whether
they would have been detected by {\it Kepler} and, if not, a new
density was sampled. The ``observational masses'' of the simulated
planets were then evaluated from the empirical piecewise power law
mass-size relation. These observational masses thus have a random
scatter compared to the input masses. Power laws of the form
$M_{p,1}/M_\oplus = p_0 r_{\rm AU}^{p_1}$ were then fit to the real
and simulated Vulcans. The real Vulcans have $p_0=7.8\pm1.5$ and
$p_1=0.72\pm 0.17$. The simulated Vulcans can achieve a similar value
of $p_0$ if $\alpha_{-3}=0.205$ (including normalization to the Paper
III disk model). They have $p_1\simeq 0.94\pm0.17$, i.e., slightly
shallower than the actual input value of $p_1=1$. Overall, this
comparison shows that the observed Vulcans have a mass versus orbital
radius relation that is consistent with the scaling predicted by gap
opening at the DZIB. The normalization of the viscosity parameter is
also consistent with theoretical expectations (Dzyurkevich et
al. 2010).

\section{Conclusions and Open Questions}

Systems with Tightly-packed Inner (Earth to Super-Earth) Planets
(STIPs) are very common. They may have formed further out in the disk
and then migrated inwards, but such models face a number of challenges
in reproducing observed architectures. Alternatively, the planets may
have formed {\it in situ}. Inside-Out Planet Formation (IOPF) is a new
{\it in situ} formation model that embraces the large mass flux of
pebbles predicted by Weidenschilling (1977) as a meter-sized barrier
for planetesimal formation. IOPF assumes these pebbles are trapped at
the pressure maximum associated with the dead zone inner boundary
(DZIB) set by thermal ionization of Na and K at $\sim 1,200$~K. A
pebble ring forms and grows to dominate the local mass surface density
and form a planet. Gap opening by this planet is the key process that
shuts off pebble accretion and leads to DZIB retreat. A new pebble
ring forms once retreat is stabilized and the process repeats.

Features of this model include that the meter-sized barrier for
planetesimal formation is not a particular problem as it is for
standard Core Accretion planet formation models. IOPF predicts planets
with masses of $\sim$few $M_\oplus$ are created on tightly-packed,
aligned orbits at distances of $\sim0.1$~AU, consistent with observed
systems. It predicts a flat scaling of planet mass with orbital
radius, again consistent with observed systems. Orbital spacings
should be at least $\sim5$ Hill radii of the inner planet, but will
typically be larger due to DZIB retreat. The Hill-normalized spacing
from first to second planet is expected to be larger than subsequent
spacings, again as observed. Innermost, ``Vulcan'' planets have a
particularly simple, linear mass versus orbital radius relation. This
is also consistent, in both its normalization and scaling, with the
observed Vulcans.

Many open questions remain to be addressed, including estimates of the
pebble supply rate to the inner disk, which sets the rate of
IOPF. Also the question of the onset of IOPF and the prior history of
the disk at earlier stages, when the accretion rates were likely
larger: in particular, why does IOPF require its onset to coincide
with inner disk accretion rates of $\sim 10^{-9}\:M_\odot\:{\rm
  yr^{-1}}$? While we expect limited migration of the first planet
when it is forming (Paper III), how much migration occurs in later
stages, including due to planet-planet interactions? Can small amounts
of H/He gas be accreted to planets forming via IOPF, i.e., under very
warm conditions? Does atmospheric evaporation play any significant
role in altering the properties of planets that may have formed by
IOPF?

If IOPF is a valid model of planet formation, why has it not occurred
in some systems, such as our Solar System? Possible reasons may
include processes that sometimes truncate pebble supply to the inner
disk, such as metallicity-dependent efficient planetesimal formation
via the streaming instability (Youdin \& Goodman 2005) leading to
early giant planet formation (e.g., Lambrechts \& Johansen
2014). Alternatively, there could be variation due to processes that
maintain ionization and MRI activity to much larger scales than
expected by thermal ionization (e.g., enhanced levels of cosmic rays
or radionuclides) or due to processes that may completely suppress the
MRI and thus remove the DZIB pressure trap in some circumstances (such
as the Hall Effect and its dependence on global disk $B$-field
orientation).

\acknowledgements JCT and SC acknowledge support from NASA ATP grant
NNX15AK20G. JCT and SM acknowledge support from a Royal Society
International Exchange grant IE131607.

\end{document}